\documentclass[conference]{IEEEtran}
\IEEEoverridecommandlockouts
\usepackage{cite}
\usepackage{amsmath,amssymb,amsfonts}
\usepackage{algorithmic}
\usepackage{graphicx}
\usepackage{textcomp}
\usepackage{xcolor}
\usepackage[hyphens]{url}
\usepackage{hyperref} 
\def\BibTeX{{\rm B\kern-.05em{\sc i\kern-.025em b}\kern-.08em
    T\kern-.1667em\lower.7ex\hbox{E}\kern-.125emX}}

\usepackage{fancyvrb}
\usepackage{mdframed}

\begin{document}

\title{A reliability measure for smart surveillance systems
}

\author{
\IEEEauthorblockN{Anj Simmons\IEEEauthorrefmark{1}}
\IEEEauthorblockA{\IEEEauthorrefmark{1}Applied Artificial Intelligence Institute, 
Deakin University, Geelong, Australia\\
Email: a.simmons@deakin.edu.au}
}


\maketitle

\begin{abstract}
We present a reliability measure for smart surveillance systems, taking into account the adversarial nature of intrusion. Our approach is based on percolation theory and is a generalisation of Hamedmoghadam et al.'s reliability measure. Specifically, our approach incorporates a customisable cost function to allow modelling a diverse range of situations, such as access restrictions, monitoring, and failures. We demonstrate our approach by applying it to a digital twin of a smart building. However, challenges remain in estimating and modelling the key parameters needed.
\end{abstract}

\begin{IEEEkeywords}
digital twins, percolation theory, reliability, surveillance, smart buildings
\end{IEEEkeywords}

\section{Introduction}

Smart surveillance systems make use of a network of sensors to automatically detect, identify and respond to intruders. For surveillance systems consisting of many sensors, ad-hoc configuration approaches becomes unmanageable, so to address this, a model (digital twin) of the home, building, or city can be used to integrate and reason about the incoming sensor data in a holistic manner rather than at an individual sensor level. While creation of digital twins was previously challenging, cloud providers now offer scalable platforms for creating such digital twins\footnote{\url{https://azure.microsoft.com/en-us/services/digital-twins/}} and integrating streams of sensor data from Internet of Things (IoT) devices.

However, economic factors limit the sensor coverage that can realistically be achieved as well as the reliability of the sensors. Even if the sensors themselves are physically reliable, power, network or AI failures can undermine the ability of the system to reliably detect and identify an intruder.

Intrusion is an adversarial scenario. If an intruder is aware that a particular system is not working they will change their behaviour accordingly to take advantage of this. For example, if they know that the AI vision model used to detect people in security camera footage is vulnerable to a physical adversarial attack (such as wearing an adversarial t-shirt \cite{Xu2020}) and they know that the software provider is yet to issue an update, they will change their attire to avoid detection by security cameras and alter their path such that they pass security cameras but avoid other kinds of sensors that they expect to still be capable of detecting them.

Furthermore, if an intruder is able to bypass one system, they will likely be able to bypass others. As an example, smart card readers may be installed on all doors and configured to deny access to unauthorised people. But if the attacker is able to obtain/forge a smart card at a certain access level (e.g. they find a staff smart card that an employee dropped), then they will be able to bypass all the smart card readers configured to permit entry at that access level.

In this paper, we present a reliability measure for smart surveillance systems, taking into account the adversarial nature of intrusion. Our approach is based on percolation theory which deals with the probability of a path emerging between two vertices in a graph or lattice, and is a generalisation of Hamedmoghadam et al.'s reliability measure, $\alpha$ \cite{Hamedmoghadam2021}.

\section{Related work}

Hamedmoghadam et al. \cite{Hamedmoghadam2021} recently proposed a reliability measure for identifying bottlenecks in transportation networks. Their approach is based on a percolation model, in which each link (edge) in the network is associated with a link quality. In the case of transport, the link quality can be defined as the current speed of that link as a fraction of the maximum speed of that link. The percolation process involves removing links from the network that do not exceed the global quality threshold, $\rho$, which is varied from 0 to 1 to study how the network connectivity deteriorates as a function of $\rho$. The key insight of Hamedmoghadam et al. is that even when global network connectivity deteriorates, sub-networks may still function effectively. To take this into account, their approach involves considering the origin destination (O-D) matrix of demand for trips between origin and desired destination, then examining what percentage of this total trip demand is possible in the network using only links that have a quality level greater than $\rho$. Specifically, the reliability measure they propose, $\alpha$, is defined as the integral (area under curve) of the percentage of trips possible with respect to $\rho$.

However, percolation analysis only focuses on whether there is a path of links connecting the origin node to the destination node, of which each must meet a minimum standard (link quality $\rho$). It has no notion of total cost, for example, travel time, which depends on the sum of travel time for each of the links making up the path (rather than the worst-case travel time within the path). This is problematic because even if a path of high quality links exists that connects the origin to the destination, the path may be impractical if it is an overly long indirect route. 
 
In our work, percolation based metrics (in which all links in the path must meet a minimum criteria) and cost based metrics (in which the combined sum along the path must meet a minimum criteria) are both specific instances of our generalised representation. In the case of surveillance, these are important for modelling access (limited by the worst-case barrier) and privacy cost (which accumulates with each device passed) respectively.

Another difference is that in transport existence of a path from origin to destination desirable; however, our interest is in restricting movement of an intruder such that no detection-free path exists. This means that in contrast to Hamedmoghadam et al. we treat a small values of $\alpha$ as desirable rather than large values.

\section{Formulation}
\label{sec:formulation}

\begin{figure}[tpb]
    \centering
    \includegraphics[width=0.7\linewidth]{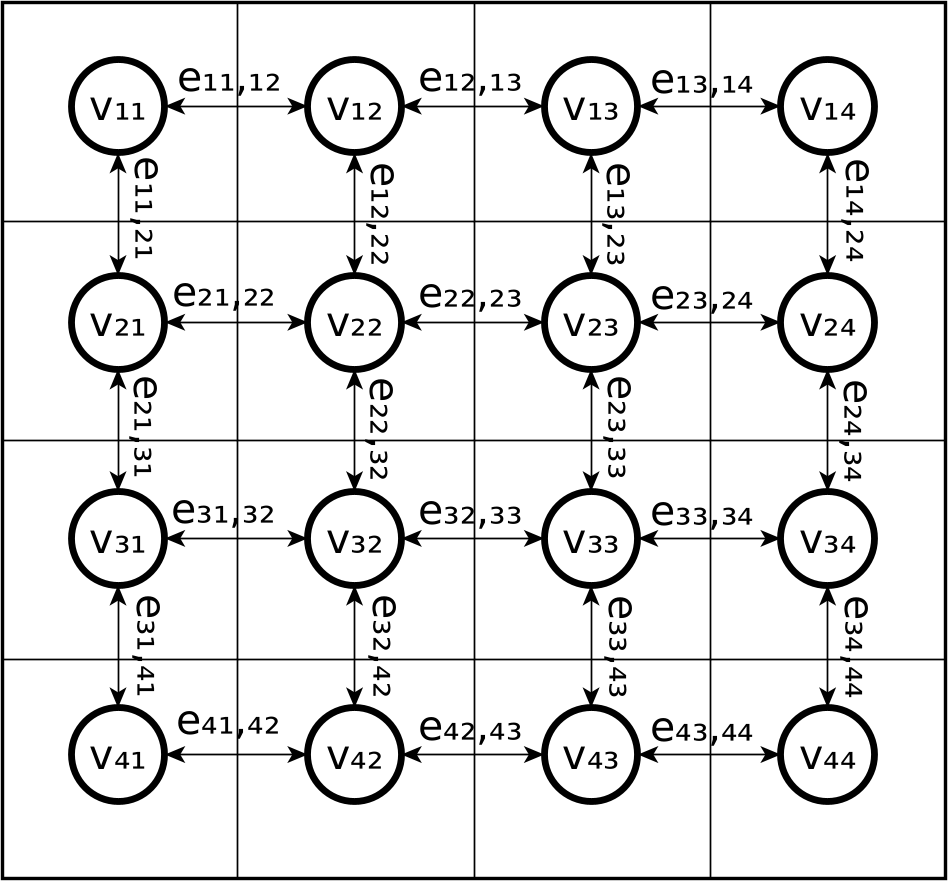}
    \caption{Two dimensional lattice, overlaid with graph representation}
    \label{fig:lattice}
\end{figure}


The traditional formulation of percolation theory\footnote{\url{https://mathworld.wolfram.com/PercolationTheory.html}} takes place on a lattice, as presented in \autoref{fig:lattice}. For ease of comparison to the graph formulation that will be used in this paper, we have labelled the set of vertices (nodes), $V = \{v_{11}, v_{12}, v_{13}, ..., v_{21}, v_{22}, v_{23}, ...\}$, and edges $E = \{e_{11,12}, e_{12,13}, e_{13,14}, ..., e_{11,21}, e_{12,22}, e_{13,23}, ...\}$ that together form the graph $G = (V, E)$ representing the connectivity of cells in the lattice.

\begin{figure}[tpb]
    \centering
    \includegraphics[width=0.7\linewidth]{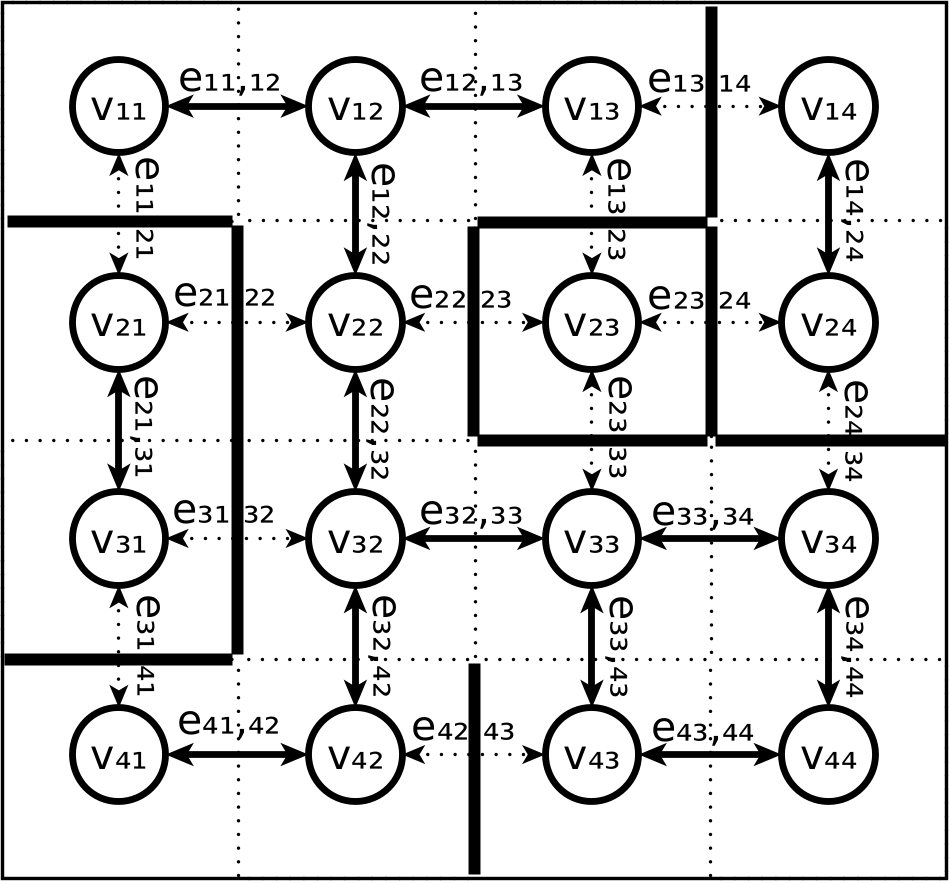}
    \caption{Example state that could arise after randomly opening/closing edges}
    \label{fig:blocked}
\end{figure}

In the Bernoulli bond percolation model\footnote{\url{https://mathworld.wolfram.com/BernoulliPercolationModel.html}}, edges are randomly either open (traversable) with probability $\rho$ or closed (blocked). An example of one possible resultant state is shown in \autoref{fig:blocked}.

In our generalised framework, we assign costs for traversing each edge. Crossing an open edge has no cost, whereas a crossing a closed edge has infinite cost. To perform the randomisation as either open or closed, for each edge we randomly sample a value $z_{ab}$ between 0 and 1 then check if this value falls below the probability of an open edge, $\rho$. The resultant cost for traversing an edge from $a$ to $b$ is presented in \autoref{eq:lattice-cost}.

\begin{align}
    \begin{split}
        c_{ab} &=
        \begin{cases}
            0, & z_{ab} < \rho \\
            \infty, & \text{otherwise}
        \end{cases} \\
    P(z_{ab}) &\sim U(0,1)
    \end{split}
    \label{eq:lattice-cost}
\end{align}

To determine if there exists a path of open edges from origin node $o$ to destination node $d$, we search for a path, $\psi$, of connected edges with minimum total cost, $c^*_{od}$. In the case of the Bernoulli bond percolation model, the resultant cost will either be zero (an open path exists) or infinite (all possible paths from the origin to the destination are blocked by at least one closed edge along the path). A theoretical formula is presented in \autoref{eq:lattice-total-cost}, although in practice this can be solved using an approach such as Dijkstra's shortest path algorithm to avoid searching an exponential number of possible paths.

\begin{equation}
    c^*_{od} = \min_{\psi \in \Psi_{od}} \sum_{e_{ij} \in \psi} c_{ij} \\
\label{eq:lattice-total-cost}
\end{equation}
Where $\Psi_{od}$ is the set of all possible paths connecting from node $o$ to node $d$ via edges in $E$.

Our general formulation allows assigning a maximum budget, $b_{od}$, for path traversal. In the case of Bernoulli bond percolation the cost will always be 0 or infinite, so $b_{od}$ can be initialised to an arbitrary positive value such as $b_{od} = 1$. We define the reachability, $r_{od}$ of the origin from the destination as cases where the cost is within budget, whereas if the cost exceeds the budget it is considered unreachable as per \autoref{eq:reachability}.

\begin{equation}
    r_{od} =
    \begin{cases}
        1 & c^*_{od} < b_{od} \\
        0 & \text{otherwise}
    \end{cases}
    \label{eq:reachability}
\end{equation}

Traditionally, percolation analysis focuses on varying the probability of edge connections, $\rho$, to find the critical threshold, $\rho_c$, at which the chance of an infinite path flowing (percolating) through an infinite lattice jumps from near zero probability to near certainty. However, when performing percolation analysis on graphs representing real transport networks, Hamedmoghadam et al. \cite{Hamedmoghadam2021} note that the critical threshold can be unclear, so instead focus on the unaffected demand that can be met by the network as $\rho$ is varied.

Following Hamedmoghadam et al. \cite{Hamedmoghadam2021}, we incorporate a term for weighting analysis by the flow demand, $f_{od}$. In the case that all origins and destinations are considered of equal importance, we can set $f_{od} = 1$ for all $o,d$. The proportion of unaffected demand, $\text{UD}_\rho$, at a given value of $\rho$ is determined by the combination of reachability and normalised flow demand, as presented in \autoref{eq:unaffected-demand}.

\begin{equation}
    \text{UD}_\rho = \frac{\sum_{o,d \in V} f_{od} r_{od}}{\sum_{o,d \in V} f_{od}}
    \label{eq:unaffected-demand}
\end{equation}

Rather than attempting to find a single percolation threshold, Hamedmoghadam et al. \cite{Hamedmoghadam2021} propose a reliability index, $\alpha$, which integrates over all possible values of $\rho$, as shown in \autoref{eq:alpha}.

\begin{equation}
     \alpha = \int_0^1 \mathbb{E}(\text{UD}_\rho) \,d\rho
     \label{eq:alpha}
\end{equation}
Where $\mathbb{E}(\cdot)$ is the expected value. The result can be computed approximately through sampling of $x_{ab}$ and $\rho$.

The key difference between our formulation of $\alpha$ and Hamedmoghadam et al.'s \cite{Hamedmoghadam2021} is that in ours traversing edges is associated with a cost function, $c_{ab}$, which can be customised to model situations other than fully open or fully closed edges. We explore the flexibility provided by the cost function within the next section. The probabilistic cost function in this section was set to be equivalent to the Bernoulli bond percolation model, and the non-probabilistic version of the cost function presented in \autoref{sec:accesslevel} is similar to the form of percolation used by Hamedmoghadam et al., whereas \autoref{sec:monitoring} and \autoref{sec:failure} express situations that have no equivalent in percolation theory. Thus, our formulation is more general than these previous definitions. However, it does not generalise all forms of percolation theory, e.g., variants formulated in a continuous space\footnote{\url{https://en.wikipedia.org/wiki/Continuum_percolation_theory}} rather than on a graph or lattice.

\section{Development of Cost Function}

The formulation presented in \autoref{sec:formulation} can serve as a general framework though specifying a custom cost function for $c_{ab}$. To illustrate this, we consider a network of possible movements by an intruder in a building. The intruder seeks to move about the building while avoiding detection, while the building designer seeks to install access checks and monitoring systems that restrict the ability of intruders to move without being detected, even if the intruder has full knowledge of the building design and its installed systems. We consider this from the perspective of access level (in which movement is restricted based on access levels), monitoring (in which intruders accumulate risk of detection, even if able to move freely about the building), and sensing failure (in which each monitoring system experiences some downtime and the intruder takes advantage of this knowledge to avoid detection).

\subsection{Access Level}
\label{sec:accesslevel}

\begin{figure}[h]
    \centering
    \includegraphics[width=\linewidth]{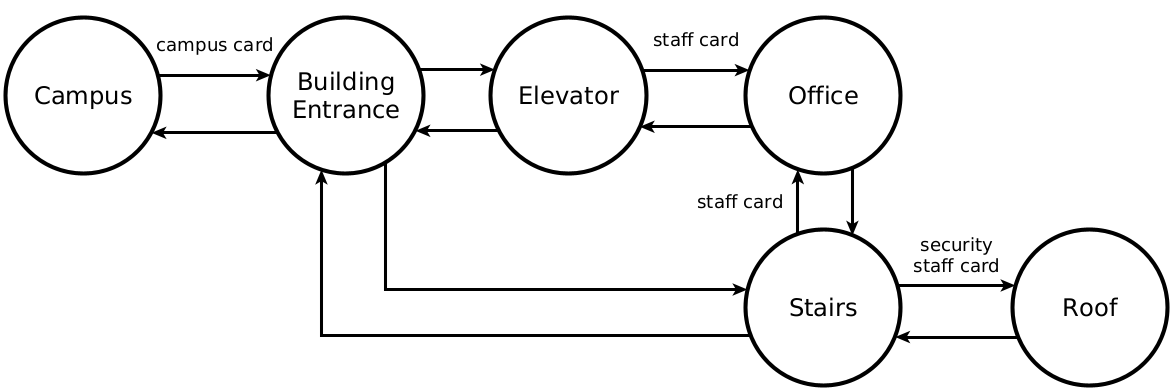}
    \caption{Access Perspective}
    \label{fig:access}
\end{figure}

\autoref{fig:access} provides an example of a university building with a staff office that can be reached via elevator or taking the stairs. Movement (edges) between different areas of the building are associated with an access level. In this scenario, the access levels are $\frac{1}{4}$ (accessible with a security staff card), $\frac{2}{4}$ (accessible with a staff card), $\frac{3}{4}$ (accessible with a campus card), and $\frac{4}{4}$ (public access). The access levels are ordered such that a security staff card grants more access than a staff card, a staff card grants more access than a campus card; and a campus card grants more access than public access.

To model this scenario, each edge is assigned a quality score, $q_{ab}$, corresponding to the access level of the edge, e.g., $q_{\text{Campus},\text{Building\,Entrance}} = \frac{3}{4}$. An intruder who has obtained/forged an access card is able to traverse any edges with a quality score above the threshold their card permits, but is blocked from traversing other edges with lower quality scores. To study the intruder's ability to move throughout the building network, we bind the intruder's access level to $\rho$, which is varied from 0 (card always granted access) to 1 (always denied access) to study how the intruder's ability to move throughout the network degrades as restrictions increase. This variation of the percolation analysis problem is similar to the approach used to study transport networks, in which edges with a quality score below $\rho$ are considered dysfunctional \cite{Li2015b,Zeng2019,Hamedmoghadam2021}.

To represent this scenario within our framework, we assign 0 penalty to links that are accessible (i.e. $q_{ab} > \rho$) and infinite penalty if not accessible. The cost function for this scenario is presented in \autoref{eq:access}. The other equations, e.g., to calculate the total cost, $c^*_{od}$, from $c_{ab}$ and to calculate the reliability metric, $\alpha$, remain the same as presented earlier in \autoref{sec:formulation}.

\begin{equation}
    c_{ab} =
    \begin{cases}
      0 & q_{ab} > \rho\\ 
      \infty & \text{otherwise}
    \end{cases}
    \label{eq:access}
\end{equation}

\subsection{Monitoring}
\label{sec:monitoring}

\begin{figure}[h]
    \centering
    \includegraphics[width=\linewidth]{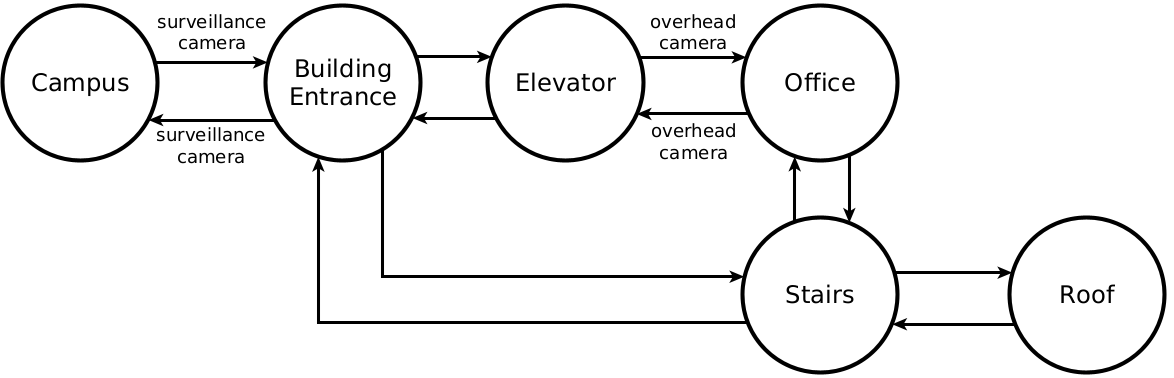}
    \caption{Monitoring Perspective}
    \label{fig:monitoring}
\end{figure}

\autoref{fig:monitoring} provides an example of monitoring systems within the building, which do not prevent intruders from movement, but instead deter intruders through increased chance of detection, particularly if the intruder passes through multiple monitored links.

Crossing a monitored edge in the building network generates some evidence that can assist in detecting the intruder, but does not in itself prove the identity of the intruder or nature of their intent unless taken together with other evidence. For this scenario, let us assume that the intruder is concerned with protecting their identity, but crossing a monitored edge reveals some information about the intruder such as their height, age, gender, or voice. The cost to the intruder's anonymity can be measured in terms of information entropy. For example, a poorly mounted blurry overhead camera that takes a top view image of the intruder's head that could match 50\% of people is obviously insufficient to identify the intruder on its own, but produces 1 bit of information which can help narrow down the intruder's identity. We will represent the expected amount of information gathered about the intruder by traversing a monitored link by $m_{ab}$, e.g. $m_{\text{Elevator},\text{Office}} = 1$ bit. Most sensors will be capable of detecting higher amounts of information, e.g. an image of the intruder taken by a better mounted surveillance camera near the building entrance may only match 1 in 1024 people, which implies $m_{\text{Campus},\text{Building Entrance}} = 10$ bits ($1024 = 2^{10}$). If both cameras take images from different views and contribute unique information, then together the total information captured from a path that passes both cameras would be 11 bits (although in practice it may be less as a result of mutual information).

A careful intruder who wishes to remain anonymous will ensure that the total amount of information they reveal is kept below their privacy budget. With just 33 bits of unique information, it is possible to uniquely identify someone amongst all other people on the planet\footnote{\url{https://www.eff.org/deeplinks/2010/01/primer-information-theory-and-privacy}}.


\autoref{eq:monitoring} incorporates the cost to the intruder from crossing links that monitor information, $m_{ab}$, about their identity. Edges that an intruder can't cross due to access restrictions ($q_{ab} > \rho$) are still treated as infinite cost (which could alternatively be thought of as a complete loss of anonymity if the intruder is caught attempting to forcefully bypass access level checks).

\begin{equation}
    c_{ab} =
    \begin{cases}
      m_{ab} & q_{ab} > \rho\\ 
      \infty & \text{otherwise}
    \end{cases}
    \label{eq:monitoring}
\end{equation}

\subsection{Failure}
\label{sec:failure}

\begin{figure}[h]
    \centering
    \includegraphics[width=\linewidth]{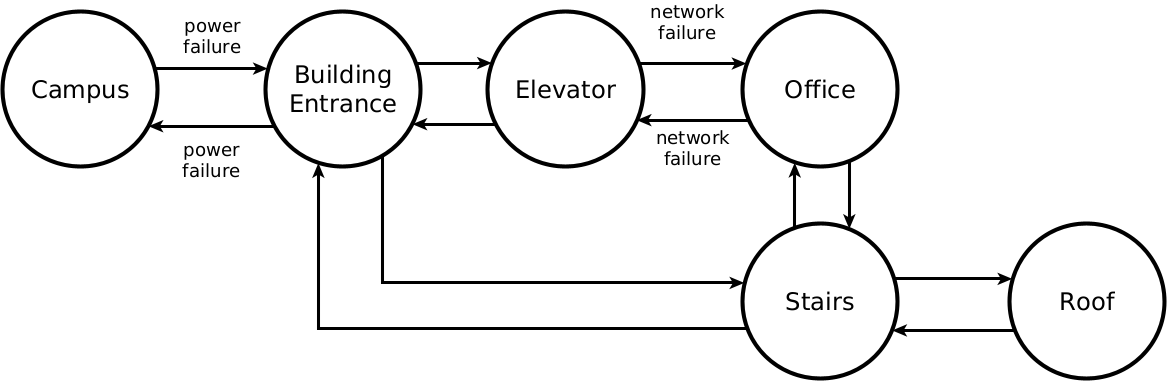}
    \caption{Failure Perspective}
    \label{fig:failure}
\end{figure}

\autoref{fig:failure} presents a scenario where random factors lead to the failure of monitoring devices. To model this, we draw a uniformly distributed random variable between 0 and 1 and check if it is below the probability of failure, $f_{ij}$. In the failed state, the sensor is assumed not to have any privacy cost to the attacker. Combining with the logic for access, we have \autoref{eq:failure}.


\begin{align}
    \begin{split}
    c_{ab} &=
    \begin{cases}
      0 & q_{ab} > \rho\ \text{and}\ z_{ab} < f_{ab} \\
      m_{ab} & q_{ab} > \rho\ \text{and}\ z_{ab} \geq f_{ab} \\ \infty & \text{otherwise}
    \end{cases} \\
    P(z_{ab}) &\sim U(0,1)
    \label{eq:failure}
    \end{split}
\end{align}

We define failure as the proportion of time that a device is possible to bypass without revealing any identifying information about the intruder. We assume that the intruder has full knowledge of the state of the building, including access levels and which monitoring devices are in a failed state. For example, they can observe from the outside of the building that there is a power failure which will cause the surveillance camera near the building entrance to stop working, and know when there is a network failure and that this will bring down the overhead camera between the elevator and office. This knowledge is incorporated into the minimisation of \autoref{eq:lattice-total-cost}. If an attacker were unaware, or the failure states were to switch so rapidly as to be unpredictable by an attacker, then sensor failure would need to be modelled differently (e.g. as a reduction of the average information collected by passing the sensors).

When a device is not in a failure state, it is available. Availability is the chance that the sensor is working at a randomly selected moment, and is equal to $1 - f_{ab}$.

A similar approach can be applied to model failure of access devices (e.g. failure of a smart card reader, which depending on whether it is configured to \textit{failopen} or \textit{failclosed}, can lead to the link to be either fully traversable or inaccessible to all).

\section{Implementation}

\begin{figure}[tpb]
    \centering
    \includegraphics[width=\linewidth]{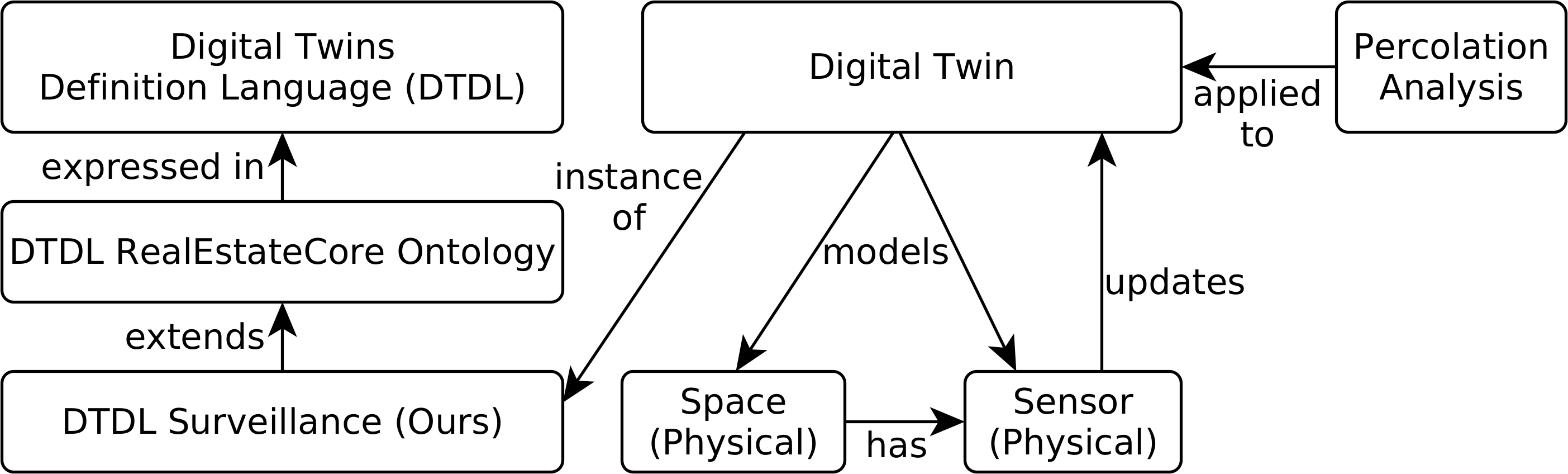}
    \caption{Our approach to representing and analysing the spaces protected by smart surveillance systems}
    \label{fig:architecture}
\end{figure}

In this section, we describe our design and implementation of a software tool to support our analysis techniques. Our tool is based on open standards with the goal of allowing it to be applied to an existing model (digital twin) of a house, building or city rather than needing to input all data from scratch. Our approach is illustrated in \autoref{fig:architecture} and explained in the following sections.

\subsection{Representing system to be analysed using open standards}

Before we can apply our analysis techniques, we first require some model of the system to be analysed. To support analysis of a wide range of systems, we build upon the concept of a digital twin, a virtual model of some physical system, e.g., of a house, building or city, including the type and placement of sensors within the system. A real-time feed of sensor observations of the system, e.g., from Internet of Things (IoT) devices can be used to update and the digital twin to ensure it is an accurate representation of the current state of the system. However, for our purposes, our interest is in the reliability of these sensors more so than the values they record.

Designing digital twins requires some form of machine readable language in which to express the model. This language needs to be expressive enough to describe all the objects in the system (the different kinds of spaces in a building, doors connecting the spaces, smart card readers used to control access to doors, security cameras, etc.) and the relationships between them. Furthermore, to stay relevant as the system changes, the language needs to be extensible to ensure that new types of objects and properties can be added after initial development. The language also needs to support interoperability as a smart building may have many different types of IoT devices from different manufacturers, and a smart city will have many different types of smart buildings. This problem lends itself to the use of ontologies to represent concepts and their relationships in an extensible manner.

To define the digital twin, we build upon the Digital Twins Definition Language (DTDL), an open standard defined by Microsoft for the purpose of defining models to represent digital twins. While not compatible with W3C standards such as RDF and OWL, it serves a similar purpose, and tools exist to automatically convert existing RDF and OWL ontologies to DTDL\footnote{\url{https://docs.microsoft.com/en-us/azure/digital-twins/concepts-ontologies-convert}}. Rather than defining a new ontology from scratch, we extend the DTDL implementation\footnote{\footnote{https://github.com/Azure/opendigitaltwins-building}} of the RealEstateCore (REC) Ontology \cite{Hammar2019} for smart buildings. There is also work towards defining a DTDL based implementation of smart cities\footnote{\url{https://github.com/Azure/opendigitaltwins-smartcities}}, but this is not as mature nor fully compatible with smart buildings, hence we focus on smart buildings as a proof-of-concept in our implementation.

Basing our work upon these open standards allows our analysis, in theory, to be applied to existing digital twins rather than requiring development of a new twin from scratch. However, our analysis also requires some additional details of sensors that aren't captured by existing standards, which we elaborate on below.

\begin{figure}[tpb]
    \centering
    \includegraphics[width=0.7\linewidth]{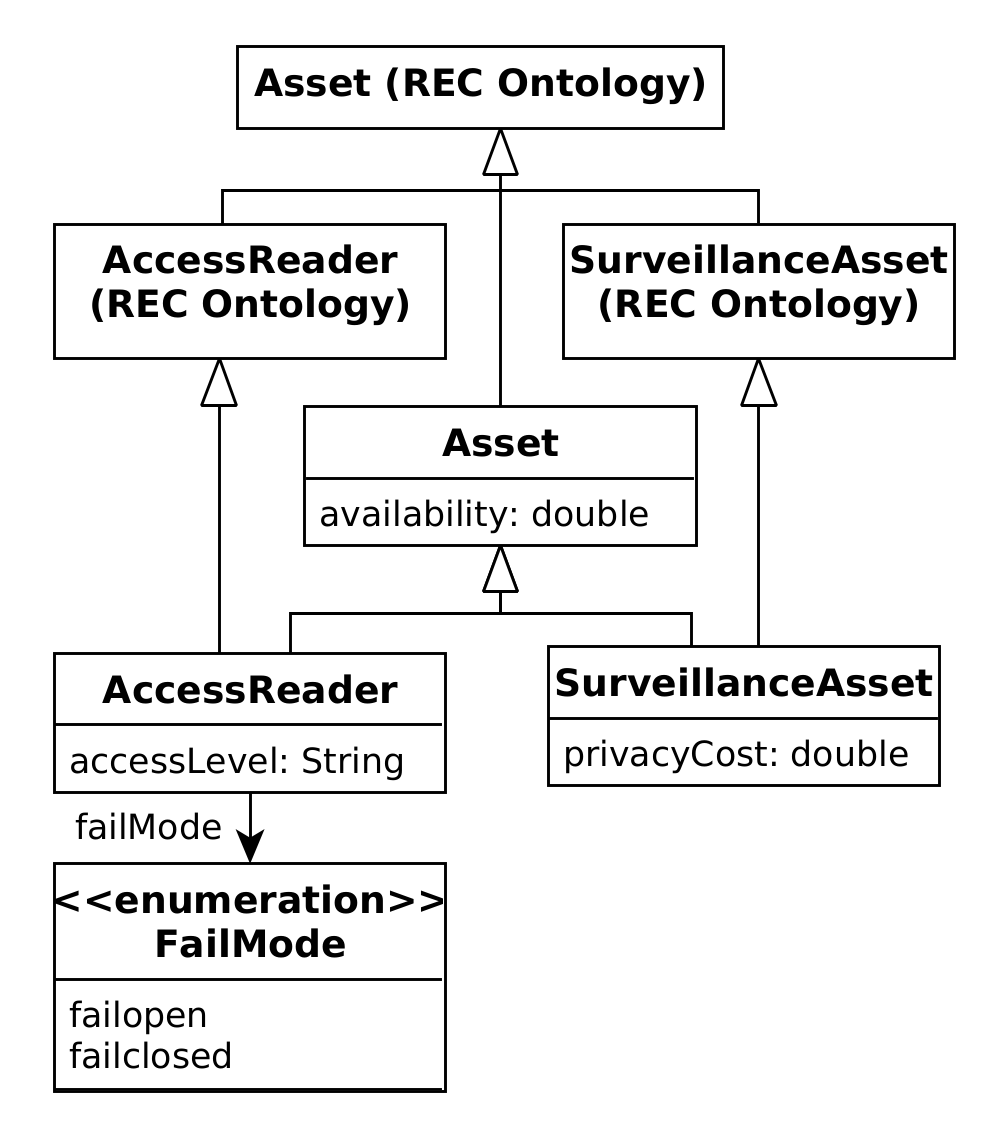}
    \caption{DTDL Surveillance Extensions}
    \label{fig:surveillance-extensions}
\end{figure}

A core part of our analysis is consideration of the reliability of sensors detecting movement in the network. To support this analysis, we extend REC with additional properties to model sensor reliability, shown in \autoref{fig:surveillance-extensions}. In particular, availability (the chance that, at a randomly selected moment, the device/sensor will be functioning correctly), privacy cost (the amount of information, measured in bits of entropy, that the device/sensor is capable of capturing about an intruder that passes by when it is functioning correctly), access level (e.g. whether a visitor, staff, or security smart card is needed) and failure mode (whether a smart card access reader will failopen or failclosed, i.e. cause the door to open or close respectively in the case of a failure). Our extensions are released publicly\footnote{\url{https://github.com/a2i2/opendigitaltwins-surveillance}} under a permissive open source licence to allow the community to use and further extend them.


\subsection{Model transformation}

For percolation analysis, we require a network representation, and not all details of the digital twin will be relevant for our analysis (as the digital twin may have been designed for reasons other than security). This requires transforming the digital twin to a simplified network representation. We developed a simple script that iterates over a digital twin graph to identify spaces to extract as nodes (e.g. rooms in a smart building), and traversable links between these nodes (e.g. doors) to extract as edges. Our script also extracts information about restrictions and monitoring along these edges, specifically, sensors (e.g. surveillance cameras) and access restrictions (e.g. smart card access readers) that doors are served by. In the case of other forms of models (e.g. in a smart city, roads and turning lanes could be analogous to building spaces and doors), a more principled approach to this task would be to make use of a model transformation language.

The extracted network is in a form suitable for directly applying the percolation analysis equations in this paper. By default, all nodes in the network are considered of equal importance, but a user may optionally specify additional information specific to the analysis, such as they are interested in movements that allow someone outside the building to traverse throughout the building and into the office without detection.

\subsection{Computational framework}

By default, we use an extended form of the cost function presented in \autoref{eq:failure} that considers both failure of links and sensors. We draw random samples, $z$, to simulate link and sensor failures. We then compute the optimal path for each origin and destination pair using Dijkstra's shortest path algorithm. We repeat the sampling process multiple times to estimate the expected value of $\text{UD}_\rho$ with respect to $z$ for a given value of $\rho$. Finally, we numerically integrate $\text{UD}_\rho$ with respect to $\rho$ to compute the reliability metric, $\alpha$, as per \autoref{eq:alpha}.

\section{Demonstration}

To demonstrate the approach, in this section we present the results of applying approach to the smart building scenario used as a running example in this paper.

The first task was to upload our surveillance extension, and its dependencies (the DTDL implementation of the REC ontology), which defines the concepts used for modelling the digital twin. Azure Digital Twins provides a model upload tool\footnote{\url{https://github.com/Azure/opendigitaltwins-tools/tree/master/ADTTools\#uploadmodels}} for this purpose.

\begin{figure}[tpb]
    \centering
    \includegraphics[width=\linewidth]{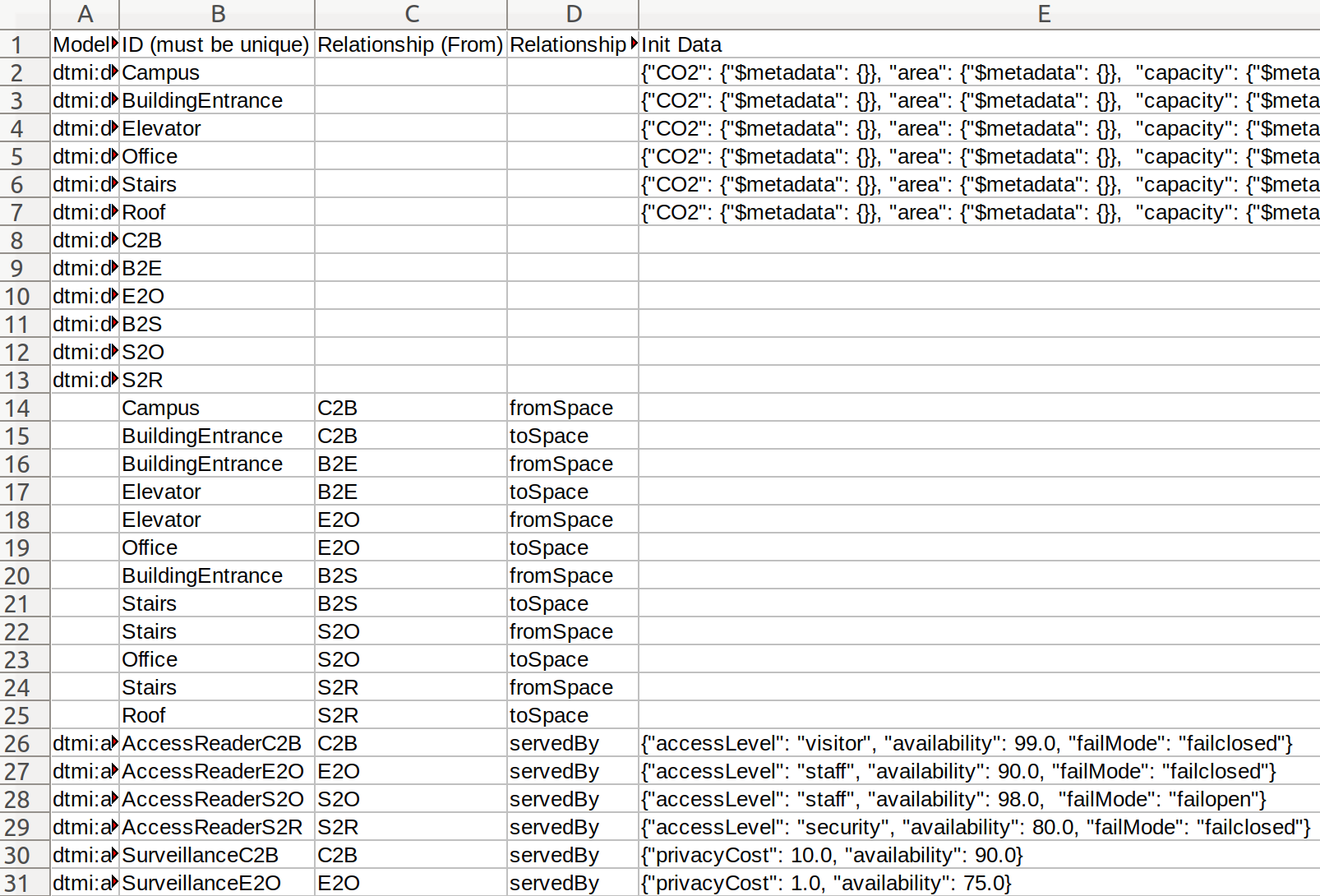}
    \caption{Spreadsheet containing digital twin data to import}
    \label{fig:spreadsheet}
\end{figure}

\begin{figure}[tpb]
    \centering
    \includegraphics[width=\linewidth]{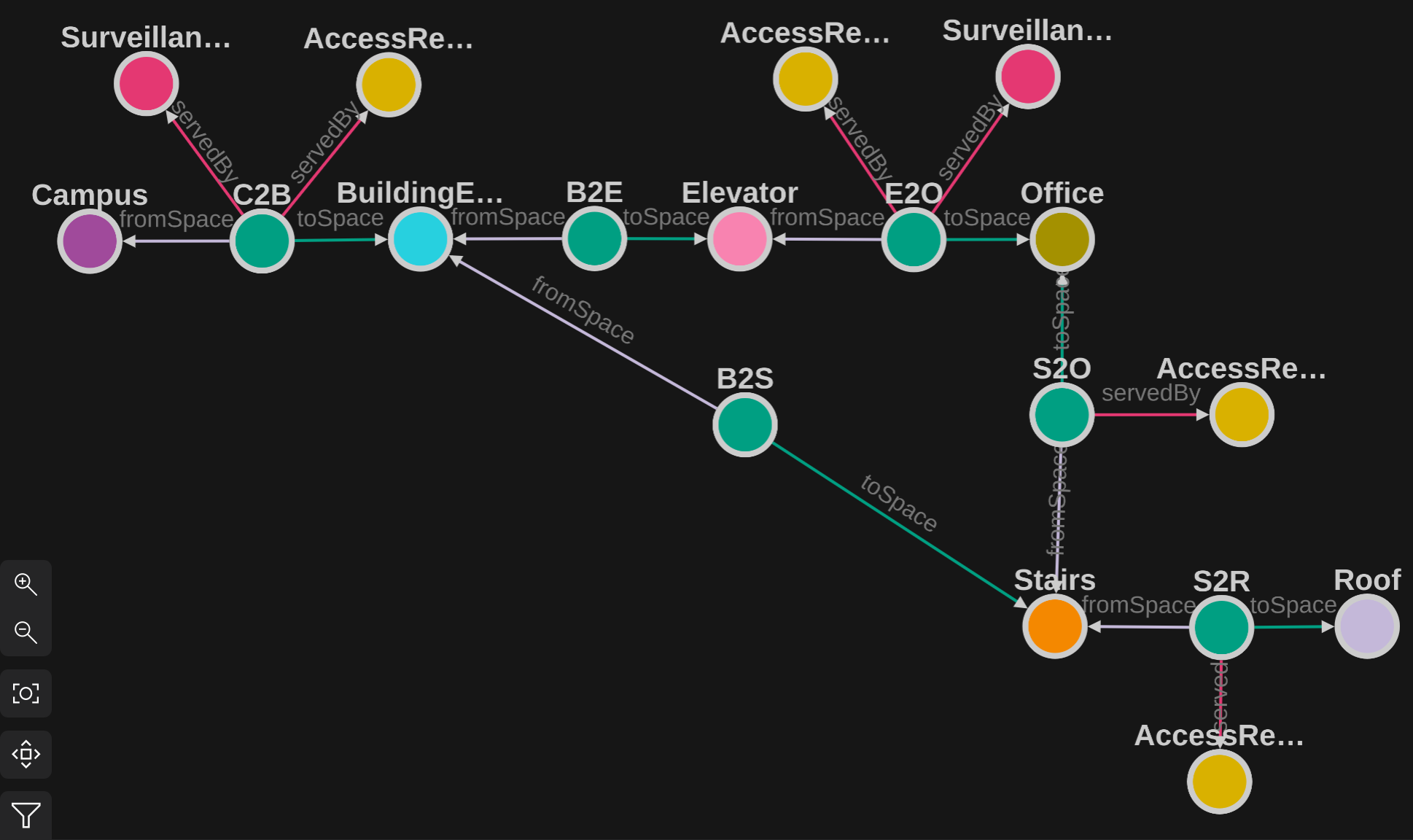}
    \caption{Digital Twin (as seen in Azure Digital Twins explorer)}
    \label{fig:twinview}
\end{figure}

The second task was to create a digital twin of the our hypothetical smart building. Data for each space, door, device and the relationships between them was specified in a spreadsheet, as shown in \autoref{fig:spreadsheet}. It is also possible to directly specify this data in JSON format (albeit, in our view, not as convenient, particularly if data needs to be sourced from stakeholders without a software engineering background). We then imported this data into Azure Digital Twins using the Azure Digital Twins explorer tool. Once imported, twin data can be updated via APIs (e.g. occupancy of a space can be updated in response to sensor measurements) and viewed/edited using Azure Digital Twins explorer. If the building had already been represented as a digital twin, then this task could have been simplified by just adding the additional properties (e.g. availability) required by our surveillance extension. The resultant twin graph is shown in \autoref{fig:twinview}.

The final step was to run our analysis scripts to perform the percolation analysis. The results are sensitive to the privacy budget, $b_{od}$, so we repeated the analysis for different values of this. An $\alpha$ value of 1 implies the intruder can move freely between all spaces in the building, whereas an $\alpha$ value of 0 implies no movement is possible without the intruder exceeding the privacy budget.

To match equation \autoref{eq:failure}, in our analysis we assumed that access card readers always function, but that surveillance devices are still subject to failure, with 99\% availability for the security camera near the building entrance (and a 10 bits of entropy when functioning), but only 75\% available for the low quality camera between the elevator and the office (and 1 bit of entropy when functioning). However, our surveillance ontology still allows specifying access availability and would be easy to incorporate into the analysis if desired. A further assumption in our analysis is that all doors are traversable in the reverse direction unless specified otherwise (without the need for an access card, e.g. so that it is possible to escape in the event of a fire) but that surveillance assets serving doors will monitor intruders passing in either direction.

We found that for our sample building, the value of $\alpha$ was (approximately) the same for all privacy budgets between 0 to 10 (note that 10 bits is the privacy cost of the surveillance camera on the building entrance), with a step increase at 10, then remains the same for all privacy budgets over 10. Results are presented in \autoref{tab:alpha-results}.

The level of surveillance between the elevator and the office turned out to be irrelevant due to the stairs. An intruder in the elevator can go down to the building entrance level, then take the stairs up to the office (or vice versa if they are in the office and wish to get to the elevator). Thus, to improve the security of the office, money would be better spent in purchasing additional surveillance devices to monitor the stairs rather than upgrading the low quality surveillance camera from the elevator to office (as an intruder with knowledge of the building plan will never need to pass this surveillance camera anyway).

\begin{table}[tpb]
    \centering
    \begin{tabular}{ c c c }
     \hline
     Budget & $\alpha$ \\ 
     \hline
     $0 < b_{od} \leq 10$ & 0.72 \\  
     $10 < b_{od} < \infty$ & 0.97 \\
     \hline
    \end{tabular}
    \vspace{1em}
    \caption{Reliability metric, $\alpha$, as a function of privacy budget, $b_{od}$}
    \label{tab:alpha-results}
\end{table}


The larger value of alpha (as the budget approaches $\infty$) is equivalent to the freedom of movement for those not concerned about being monitored (i.e. constrained only by access level and faulty smart card readers). This is important to consider, as good security should limit movements of an intruder, while still ensuring freedom of movement for others. For example, an alternative solution to improve security is to change the door between the stairs and office to a one way door that allows people to use it to exit the office but not to enter, so as to fix the issue we observed where the attacker uses the stairs as a means to bypass the surveillance camera between the elevator and office. However, doing so would also result in a change of freedom of movement even for those not concerned about being monitored, making it impossible to get to the office on days when the access card reader between the elevator and office is faulty and would make it harder for firefighters to access the office to extinguish a fire.

\section{Challenges}

When reusing an existing digital twin, it will still be necessary to specify the availability, privacy cost, access level, etc. required by our surveillance extensions in order to support the analysis. While we expect this to be preferable to redesigning a digital twin from scratch, it will take time and consideration to determine these values, and in some cases they may be unknown.

\subsection{Modelling Access Level}

Access levels are more objective than the other estimates we require, and some access control software providers allow this information to be exported or accessed via an API\footnote{\url{https://api.kisi.io/docs\#/operations/fetchGroupLocks}}. However, a limitation of our current approach is the assumption that there are ordered access levels (visitor, staff, security) rather than specialised groups which don't necessarily have any more or less permissions relative to other groups (e.g. separate access groups for each building level). In this situation the mapping of categorical groups to the quality score, $q_{ab}$, poses a problem. This requires allowing categorical values of $q_{ab}$, and then calculating $\alpha$ by summing the expected unaffected demand for each group (or permutation of groups in cases that a single person can be assigned to multiple groups) rather than integrating as a function of a single real variable.

An assumption of our approach is that all reachable spaces are connected by some form of door. However, if the digital twin was created for a different purpose, it is possible that it will not include all doors between spaces, e.g. it may be implicit that spaces on the same building level are connected. Furthermore, an intruder may find ways to move between rooms in ways that the designers never intended, e.g., breaking through windows or climbing walls. This becomes even more challenging at a city level where outdoor spaces provide even further options for an intruder such as tunnelling under fences, swimming lakes or flying a helicopter onto a roof. Our approach is thus more suitable for analysing cases in which possible movements are well understood and an intruder is not willing to risk the attention or legal penalties of resorting to physical damage or overtly suspicious behaviour.

\subsection{Estimating Privacy Cost}

Our approach assumes that the privacy cost to the intruder can be measured in terms of information entropy, and that this accumulates every time the attacker passes a sensor. However, if some information collected about the attacker is correlated with previous information, than the new information gained about the attacker will be less than expected. For example, 20 minutes of vision footage all from the same angle is not as informative as 5 minutes of vision footage from different angles along with a voice recording. On the other hand, if there is no mutual information, then it will difficult to determine that data collected from different sensors relates to observations of the same individual. Furthermore, information gained is culturally and temporally dependant rather than a constant. For example, in mid-2020 it became common for countries to recommend wearing face masks, but a side effect of this is reducing the information that can be gained via facial recognition and interference with AI models.\footnote{Apple iPhone users were unable to unlock their phone while wearing a mask until early 2022 due to the difficulty of extracting enough information to distinguish their face from that of an intruder. \url{https://www.theverge.com/2022/2/2/22912677/apple-face-id-mask-update-ios-15-4-beta-hands-on-impressions}}


\subsection{Estimating availability}

To prevent the need to individually estimate the availability, these values could be computed empirically. The availability of a sensor, which we define as the chance that it is working at any randomly selected moment, is the same as the percentage uptime, $\frac{\text{total uptime}}{\text{total uptime} + \text{total downtime}}$. Server logs could be used to detect downtime where power, network or software failures prevent the sensor from reporting reporting values. However, we also need to include cases where the sensor is reporting values but malfunctioning as downtime. This could be achieved through detecting deviations from typical patterns, comparison to other nearby sensors, and predefined rules to detect known failure states. Finally, we also need to incorporate cases where everything appears to be working but the AI is vulnerable to a physical adversarial attack that allows an intruder to evade detection as downtime. This is perhaps the hardest to estimate, as unlike the others, such a vulnerability is invisible until exploited. While the AI community closely tracks the average performance of computer vision algorithms against benchmark datasets, there are limited benchmarks to track the worst-case performance of models against adversarial attacks\footnote{\url{https://paperswithcode.com/task/adversarial-attack}}, particularly physical adversarial attacks taking place in the real-world. What is needed is a platform to upload videos featuring an attempted physical adversarial attack, and to track the worst-case performance of AI models against these. AI service providers could then issue an alert when an easy to reproduce attack is found in an AI model used for surveillance purposes, and advise to switch to a different AI model.

A further limitation of our approach is the assumption that each device fails independently, which is unlikely in practice. For example, a power or network fault could cause all devices on a floor, or even entire building, to simultaneously fail. This would provide a window of opportunity for an intruder which is unlikely to occur if they had to wait for every sensor along a path to fail independently. To model such situations, we plan to explore approaches based on fault tree analysis, for example, to model that monitoring devices are dependent on shared power, networks, and AI services. Furthermore, due to the adversarial nature of an intrusion, the intruder may deliberately cause these underlying services to fail, e.g. by jamming wireless networks, cutting power, or cyberattacks.

\section{Conclusions}

In this paper, we showed the link between percolation theory and the reliability with which a smart surveillance system will be able to detect an intruder. Furthermore, we demonstrated how our approach can be practically applied to a digital twin of a smart building. Future work is needed to more accurately model access, monitoring and failures as well as practical methods to estimate parameters. Our reliability metric is intended to help guide efficient decision making to improve security, but is unlikely to provide meaningful comparisons between different buildings.

\section{Broader Impact}

An obvious next step is applying our approach to model the reliability of surveillance systems for smart cities. While our approach already incorporates many of the key concepts needed to describe a city (buildings, spaces, and `doors' between these spaces), modelling an attack on a city is more challenging than an intrusion of a building due to the greater flexibility an attacker has. At city scale, algorithm efficiency also becomes more important, for example it would be necessary to replace our implementation of Dijkstra's algorithm used to solve \autoref{eq:lattice-total-cost} with a more efficient approximate solution.



\section*{Acknowledgment}

This paper was supported by research funding from the National Intelligence Postdoctoral Grant program (NIPG-2021-006).

\bibliographystyle{IEEEtran}
\bibliography{refs}

\end{document}